# Synthesis, Crystal Structure and Properties of a Perovskite-Related Bismuth Phase, $(NH_4)_3Bi_2I_9$

Shijing Sun,[a] Satoshi Tominaka,[b] Jung-Hoon Lee,[a] Fei Xie,[a] Paul D. Bristowe[a] and Anthony K. Cheetham[a*]

[a]Functional Inorganic and Hybrid Materials Group, Department of Materials Science and Metallurgy, University of Cambridge, U.K.
[b]National Institute for Materials Science, Tsukuba, Japan

**Abstract**

Organic-inorganic halide perovskites, especially methylammonium lead halide, have recently led to a remarkable breakthrough in photovoltaic devices. However, due to the environmental and stability concerns of the heavy metal, lead, in these perovskite based solar cells, research in the non-lead perovskite structures have been attracting increasing attention. In this study, a layered perovskite-like architecture, $(NH_4)_3Bi_2I_9$, was prepared in solution and the structure was solved by single crystal X-ray diffraction. The results from DFT calculations showed the significant lone pair effect of the bismuth ion and the band gap was measured as around 2.04 eV, which is lower than the band gap of $CH_3NH_3PbBr_3$. Conductivity measurement was also performed to examine the potential in the applications as an alternative to the lead containing perovskites.

**Introduction**

Renewable energy sources, such as solar, hydroelectric and wind power has been receiving increasing attention in the 21st century in order to meet the growing global energy demand as well as the environmental considerations. Not surprisingly therefore research and development in promising new materials and devices that improve the use of renewable energy, for example solar cells and eco-friendly light emitting diodes, have been under the spotlight in both academia and industry. Among these, a relatively young class of hybrid materials, namely organo-lead trihalide perovskites has recently struck the solar cell community with its outstanding performances in photovoltaic devices. In particular, the perovskite based solar cells using methylammonium lead iodide ($CH_3NH_3PbI_3$) have achieved the soaring power conversion efficiency from just over 3% in 2009[1] to over 20% by the end of 2014.[2] Represented by $CH_3NH_3PbI_3$, this family of organic- inorganic perovskites offer a number of advantages such as large absorption coefficients and high charge carrier motilities, as well as the ability of depositions via scalable solution processing at relative low temperatures, in comparison with its counterparts based on inorganic compounds.[3,4] However, the commercialisation of organo-lead halide materials also faces a number of challenges. One of the major concerns of these materials is the presence of the heavy metal, lead, which is toxic in addition to its instability in moist air.[5,6] Snaith et.al and Kanatzidis et.al have reported to replace Pb with other Group IV metals, such as Sn[7,8] and Ge.[9] However, the preferred oxidation state of $Sn^{4+}$ and $Ge^{4+}$ reduce the stability of these three dimensional perovskite structures in air.

Beyond the Group IV metal, our attention is drawn to the opportunities to use bismuth, which is chemically similar but more environmentally friendly, as an alternative to lead in the organometallic halide perovskite architecture. Bismuth has been used in medicine, for example, bismuth subsalicylate (under brand name of Pepto-Bismol) is available in the form of chewable tablets and is used to treat occasional upset stomach, heartburn, and nausea as well as diarrhea.[10] The conventional three dimensional hybrid perovskite consists of $ABX_3$ formula (where A is a monovalent cation, B is a divalent metal and X is a halide ion) and hence it is challenging to accommodate a $Bi^{3+}$ instead of a $Pb^{2+}$ ion at the A site. The replacement of $PbI_2$ with $BiI_3$ with the same synthesis condition as $CH_3NH_3PbI_3$ we obtained methylammonium bismuth iodide, $(CH_3NH_3)_3Bi_2I_9$, which is a 0D dimer at the room temperature (See Figure S5 and Table S2), isothermal to $(CN_3H_6)_3Sb_2I_9$ and $(CN_3H_6)_3Bi_2I_9$ reported by Szklarz et al. [11], as well as $Cs_3Bi_2I_9$ at room temperature.[12] This is not surprising, since organic-inorganic hybrid materials process the great advantages of the structural diversity by tuning the compositions.[13] As studied extensively by Mitzi et al,[14,15] the dimensionality of the perovskites varies from 0D to 3D depending on the sizes of the ions, forming isolated polyhedral, chains, layers or three dimensional perovskites. Recently layered perovskites have been reported in solar cells devices, demonstrating the band gap tuning by changing the size of cations and the dimensionality of the framework. For example, devices using the layered perovskites, $(PEA)_2(MA)_2[Pb_3I_{10}]$ (band gap of 2.06 eV) have achieved a power conversion efficiency of 4.73%,[16] and $(CH_3(CH_2)_3NH_3)_2(CH_3NH_3)_{n-1}Pb_nI_{3n+1}$ (n = 1, 2, 3, and 4) (from band gap of 1.52 eV for n = ∞ to 2.24 eV for n =1 ) leads to an efficiency of 4.03% for n=3.[17] The applications of these wide band-gap layered perovskite materials encouraged us to explore the layered perovskites with structure of $ABX_4$, where in this case A is a monovalent cation, B is a trivalent metal and X is a monovalent anion. [18]

With the attempt to synthesise the layered perovskite structure isostructural to $NH_4FeF_4$ [19], a 2D layered perovskite-related architecture, $(NH_4)_3Bi_2I_9$ was obtained, which is isostructural to the hexagonal $Rb_3Bi_2I_9$[20]. However, the ammonium phase was solution processable, in comparison to other alkaline metal based layered perovskites that were previously reported forming by solid state reactions. Furthermore, this structure type is also known in the Sb based compounds, such as $(NH_4)_3Sb_2I_9$[21] and $Rb_3/Cs_3Sb_2I_9$.[22] In their recent publication, Mitzi et al characterized $Cs_3Sb_2I_9$ thin-film [23] and reported the potential of this family of hybrid materials for photovoltaic applications.

In this study, we report the crystal structure and physical properties of $(NH_4)_3Bi_2I_9$, which can be harvested in solution from the same experimental conditions as $CH_3NH_3PbI_3$. The experimental obtained structures were compared with DFT calculations and thermal analysis were carried out to assess the stability. Conductivity and optical measurement were also performed to examine the potential in the applications as an alternative to the lead containing perovskites.

**Methods**

1) Synthesis

Single crystals of $(NH_4)_3Bi_2I_9$ were prepared via solution method. 1 mmol of bismuth iodide (Sigma-Aldrich) was dissolved in 0.3 mL of hydroiodic acid (57% aqueous solution, Sigma-Aldich) up on heating in an oil bath. At 90 °C, stoichiometric amount of ammonium iodide (Sigma-Aldrich) was added to the solution and the mixture was then kept still for 3 hours at this temperature. Dark red crystals (see Figure 2) were then immediately filtered out without leaving the solution to cool to room temperature. The crystals were dried under vacuum overnight. Red fine powder was obtained by grinding at room temperature.

2) X-ray Diffraction

Powder X-ray diffraction was performed at room temperature to confirm the phase of a ground bulk sample (see Figure S1). Single crystal X-ray diffraction was carried out using an Oxford Diffraction Gemini E Ultra diffractometer with an Eos CCD detector. Mo Kα radiation (λ = 0.7107 Å) was applied. Temperature range of 120 K to 380 K was set up under nitrogen flow. Data collection and data reduction were performed with CrysAlisPro (Agilent Technologies). An empirical absorption correction was applied and using the Olex2 platform, [24] the structure was solved with ShelXT[25] using direct methods and refined with ShelXL[26] by a least squares minimization algorithm. All non-hydrogen atoms have been refined anisotropically. A summary of the crystal structure information is shown in Table 1.

3) Materials Characterization

UV spectrometry was carried by a PerkinElmer Lambda 750 UV-Visible spectrometer. The measurement was in optical reflectance mode, with 1nm slit width. The scan interval is 1 nm and scan range covers the absorption edge. Samples were grinded into fine powder, with powder size of around 10 μm and thickness around 1 mm. Band gap was calculated using Tauc relation (see Figure S2).

Thermogravimetric analysis (TGA) was carried out under continuous nitrogen flow using a TA Instruments Q-500 series thermal gravimetric analyser. 6.31 mg of the sample was held on a platinum pan. The sample was heated at a rate of 10 °C min$^{-1}$ upto 800 °C and the significant weight loss was observed starting from around 240 °C (see Figure S3).

CHN analysis was performed to confirm the elemental composition, the results show H: 0.80%, N: 2.5% (H: 0.74%, N: 2.6% by calculations).

4) DFT Calculations

All the DFT calculations employed the generalized gradient approximation (GGA) implemented with projector augmented-wave (PAW)[27,28] pseudopotentials as supplied in the Vienna ab initio Simulation Package (VASP)[29–32] The spin-orbit coupling effect was only included during band structure and density of state calculations. The effects of van der Waals dispersion interactions were included during structural and electronic relaxation.[33] The following parameters were adopted: (i) a 3×5×2 Monkhorst-Pack k-point mesh for the layered structure. [34] (ii) a 500 eV plane-wave cutoff energy. The number of valence electrons treated explicitly were as follows: 15 for Bi ($5d^{10}6s^26p^3$), 7 for I ($5s^25p^5$), 6 for O ($2s^22p^4$), 5 for N ($2s^22p^3$), and 1 for H ($1s^1$). All structural relaxations were performed with a Gaussian broadening of 0.05 eV.[35] The ions were relaxed until the forces on them were less than 0.01 eV·A$^{-1}$. All schematic representations of the crystal structures were generated using the VESTA program.[36]

5) Conductivity Measurement

The conductivities of the single crystals were measured using the single crystal conductivity method.[37,38] In brief, single crystal samples were filtered and dried in the vacuum oven overnight. The dried crystals were mounted on two Au microelectrodes which were designed to have a 20 micro meter gap by lithography. The electrode was placed in a dark chamber with dry nitrogen flow. The conductivities were measured by the AC impedance method using the Gamry Interface electrochemical instrument at 10 mV amplitude from 1 MHz to 0.1 Hz at 20°C. We also performed powder conductivity measurements on a pellet sample. The pellet was formed at 0.1 GPa for dry powder, and then, before measuring conductivities, the pellet was kept in dry nitrogen for a week, in order to remove adsorbed water which caused ionic conduction probably due to hydrated interparticle spaces.[39]

## Results and Discussions

$(NH_4)_3Bi_2I_9$ crystallises in a monoclinic system with space group of $P2_1/n$ and lattice parameters of a = 14.6095(4) Å, b = 8.1427(3) Å, c = 20.9094(5) Å and β = 90.917(2)° (See Table 1). The framework of $(NH_4)_3Bi_2I_9$ consists of $BiI_3$ layers stacked in a closed-packed fashion, forming hexagonal layers along the *ab* plane, as shown in Figure 1 (a) and (b). Each $Bi^{3+}$ atoms is coordinated by six I-ions in a distorted octahedral environment, with Bi –I bond distances vary between 2.93 Å and 3.27 Å (three closer and three further away). The $BiI_6$ octahedra are connected by sharing corners and there are Bi atoms filled in two-thirds of the available $BiI_6$ octahedra cavities. As seen in Figure 1 (c), the structure of $(NH_4)_3Bi_2I_9$ is hence related to the conventional 3D perovskites, as it is the structure obtained after removing every third layer of the octahedral along the <111> direction of a cubic $ABX_3$ perovskite.

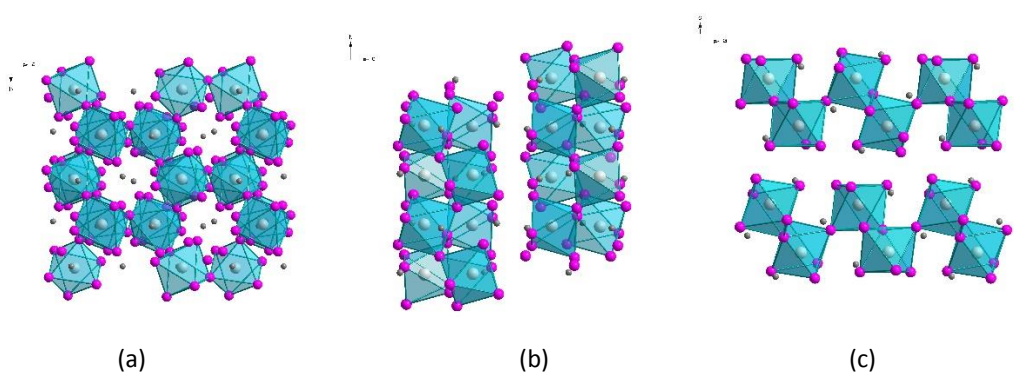

(a)　　　　　　　　　　　(b)　　　　　　　　　　　(c)

**Figure 1** Crystal structure of $(NH_4)_3Bi_2I_9$ phase along (a) *c*-axis, (b) *a*-axis and (c) *b*-axis. The silver spheres represent bismuth ions, purple spheres represent iodide ions and grey spheres represent nitrogen in $NH_4$ ions.

**Table 1** Crystallographic data and refinement of $(NH_4)_3Bi_2I_9$.

| Empirical formula | $N_3H_{12}Bi_2I_9$ |
|---|---|
| Temperature/K | 300 |
| Crystal system | Monoclinic |
| Space group | $P2_1/n$ |
| a/Å | 14.6095(4) |
| b/Å | 8.1427(3) |
| c/Å | 20.9094(5) |
| α/° | 90 |
| β/° | 90.917(2) |
| γ/° | 90 |
| Volume/Å³ | 2487.06(12) |
| Z | 4 |
| $\rho_{calc}$g/cm³ | 4.312 |
| μ/mm⁻¹ | 25.29 |
| F(000) | 2656 |
| Crystal size/mm³ | 0.20 × 0.17 × 0.11 |
| Goodness-of-fit on $F^2$ | 1.001 |
| Final R indexes [I>=2σ (I)] | $R_1$ = 0.0362, $wR_2$ = 0.0710 |

Unlike $Cs_3Sb_2I_9$, which shows the polymorphism of 0D dimer and 2D layered structures,[23] this phase of $(NH_4)_3Bi_2I_9$ is dark red in colour and the results from variable temperature single crystal X-ray diffraction reviews no evidence of phase transition and the layered phase is stable across the temperature range between 120 K to 380 K. The thermal expansion of the cell is shown in Figure 2 (b) (See Figure S4 and Table S1 for volume change and coefficients of thermal expansion).

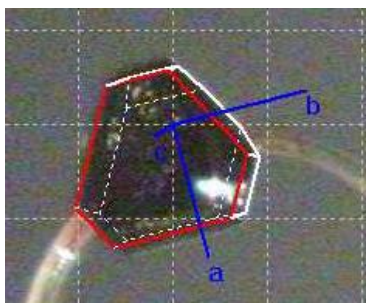

(a)

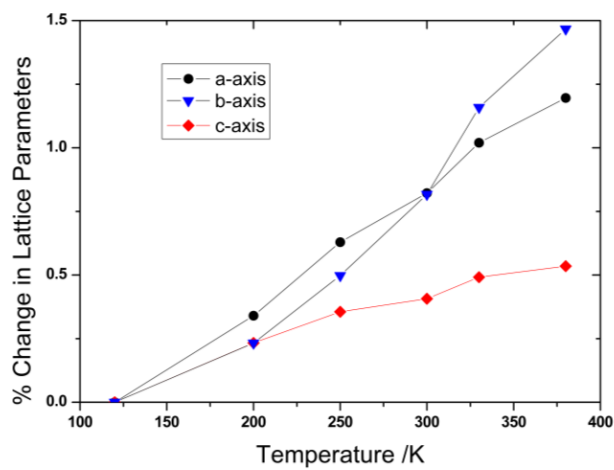

(b)

**Figure 2** (a) Photograph of (b) Change in lattice parameters as a function of temperature from 120 K to 380 K.

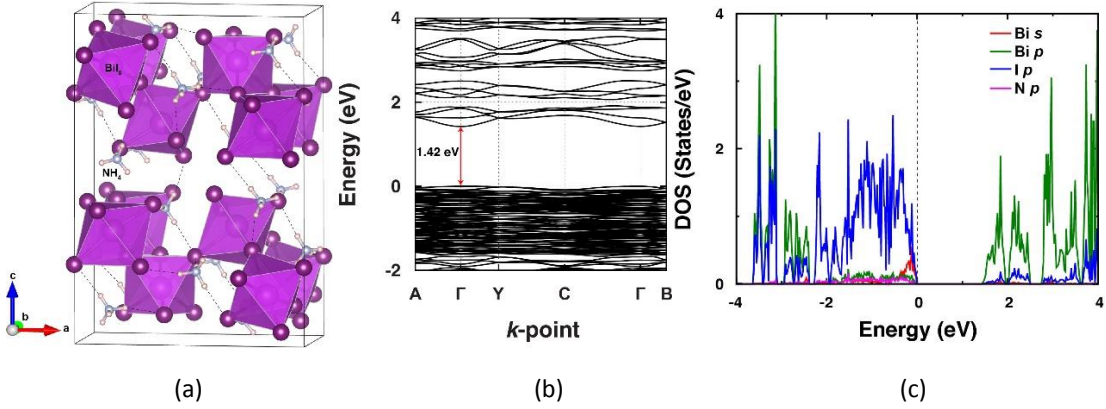

(a) (b) (c)

**Figure 3** (a) Optimized crystal structures of the layered (NH$_4$)$_3$Bi$_2$I$_9$, the dotted lines indicate hydrogen-bonds between I and H and O and H atoms. (b) Band structures along the *k* vectors of the high symmetry line of the first Brillouin zone. (c) Partial density of states for (a) Bi *s*, Bi *p*, I *p*, and N *p* orbitals.

DFT calculations were performed and it accomplished the experimental study. The following ground-state lattice parameters at zero pressure were obtained from DFT calculations: the unit-cell parameters of *a* = 14.5029 Å, *b* = 8.1984 Å, *c* = 20.6732 Å, and *β* = 90.14°. These values are in good agreement with our experimental values. The optimized atomic positions are also listed in Table 2. The computed atomic positions agree very well with our experimental results. In particular, we have successfully predicted the preferred positions of the H atoms, which is hard to determine via experimental X-ray diffraction, due to the low electron density of H atoms in comparison with Bi and I atoms. In fact, the positions of the H atoms conform with the *P2$_1$/n* space group. In Figure 3(a), hydrogen-bonds are indicated by the dotted lines. We assume that it is regarded as hydrogen-bond when H --- I and H --- O < 3 Å. Interestingly, the hydrogen-bond interaction plays a key role in stabilising the layered. Namely, hydrogen-bonds provide links between the BiI$_6$ octahedra and thereby stabilize the whole structure.

**Table 2** Computed lattice parameters and atomic positions of the *P2$_1$/n* (NH$_4$)$_3$BiI$_9$ structure.

| | | | |
|---|---|---|---|
| *a* | 14.5029 Å | | |
| *b* | 8.1984 Å | | |
| *c* | 20.6732 Å | | |
| *β* | 90.14° | | |
| | *x* | *y* | *z* |
| Bi1 | 0.6633 | 0.5117 | 0.3456 |
| Bi2 | 0.3324 | 0.5028 | 0.1555 |
| | | | |
| I1 | 0.7580 | 0.1858 | 0.2840 |
| I2 | 0.3739 | 0.1943 | 0.0849 |
| I3 | 0.4605 | 0.7019 | 0.0756 |
| I4 | 0.3027 | 0.8216 | 0.2436 |
| I5 | 0.5705 | 0.8176 | 0.3858 |
| I6 | 0.5837 | 0.3047 | 0.4490 |
| I7 | 0.1650 | 0.5779 | 0.0764 |
| I8 | 0.8375 | 0.5729 | 0.4202 |
| I9 | 0.4929 | 0.4042 | 0.2559 |
| N1 | 0.5313 | 0.9785 | 0.2240 |
| N2 | 0.8131 | 0.0082 | 0.4379 |
| N3 | 0.3503 | 0.5143 | 0.3966 |
| H1 | 0.5688 | 0.4707 | 0.1813 |
| H2 | 0.9768 | 0.1013 | 0.2642 |
| H3 | 0.5653 | 0.4197 | 0.2617 |
| H4 | 0.7861 | 0.5597 | 0.3958 |
| H5 | 0.7671 | 0.5213 | 0.4761 |
| H6 | 0.8757 | 0.5641 | 0.4485 |
| H7 | 0.3890 | 0.0281 | 0.3548 |
| H8 | 0.2860 | 0.0701 | 0.3915 |
| H9 | 0.3240 | 0.1163 | 0.9303 |
| H10 | 0.6158 | 0.9341 | 0.5639 |
| H11 | 0.6588 | 0.1103 | 0.5950 |
| H12 | 0.4663 | 0.4264 | 0.2183 |

As shown in Figure 3(b), the computed band gaps of the layered are 1.42 eV and it has a direct band gap. Our DFT calculations give the quiet flat valence and conduction bands. It is interpreted as the heavy effective mass of electrons and holes. Partial density of states (PDOS) for various atomic orbitals was then examined. In Figure 3(c), we compute the orbital-resolved PDOS for Bi *s*, Bi *p*, I *p*, and N *p* of the layered structure. The two prominent features in the PDOS are: (i) a weak overlapping of the Bi 6*s* orbital PDOS with the I 5*p* orbital PDOS for an energy range between -0.5 and 0 eV below the valence-band top and (ii) a overlapping of the Bi 6*p* orbital PDOS with the I 5*p* orbital PDOS for an energy range between 1.5 and 4 eV above the conduction-band minimum. These orbital levels can further be visualized in real space by presenting the isosurface plots of computed partial-charge density (PCD) of the topmost band below the valence-band top and the bottommost band above the conduction-band minimum. Figure 4 shows the highest occupied molecular orbital (HOMO) is mainly composed of Bi 6s – I 5p antibonding state, and the lowest unoccupied molecular orbital (LUMO) is mainly composed of Bi 6p – I 5p antibonding state in common. Interestingly, the HOMOs clearly show the stereo-active Bi 6s lone-pair orbitals as shown in Fig. 4(a). Because of these lone-pair electrons, Bi ions are displaced with respect to the center of $BiI_6$ octahedra (See Figure S7 and Table S3-5 for experimental bond lengths and angles).

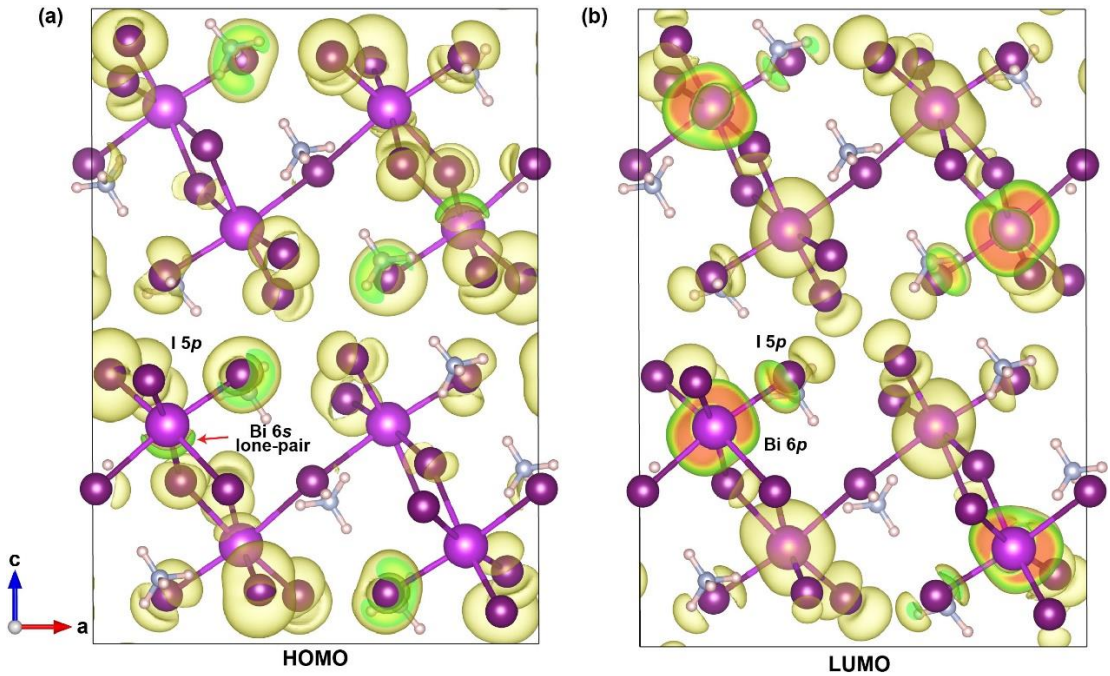

**Figure 4** Isosurface plots of the PCD of (a) the HOMOs and (b) the LUMOs of layered $(NH_4)_3Bi_2I_9$. The isosurface level is equal to 0.0008 e/Å$^3$.

Experimentally, band gap of the powder sample was measured as 2.04 eV using Tauc relation, (See Figure S2), and this is lower than the band gap of the 0D dimer bismuth phase $CH_3NH_3)_3Bi_2I_9$, which was measured as 2.16 eV (See Figure S5). The band gap of $(NH_4)_3Bi_2I_9$ with layered structure is also lower than the 3D perovskite, $CH_3NH_3PbBr_3$, which was reported with band gap of 2.20 eV.[40]

In order to further examine the potential of $(NH_4)_3Bi_2I_9$ as a photovoltaic material, measurements on conductivity was performed. Several pieces of crystals were tested, and we confirmed that they were, at least intrinsically, insulating judging by high resistivity (>3.1 x 10$^8$ Ωm at 20°C measured along a diagonal line of a hexagonal plate (*ab* plane), 35 μm thick and 178 μm wide, mounted on 20 μm gap)[note]. Considering the fact that the analogous structures have similar band gaps as our compounds and can work as the absorption layer for photovoltaics, these compounds might be extrinsically conductive. In fact, the AC impedance data for the powder sample shows a low resistivity of 0.42 Ωm (20°C) without phase shifts below 10 kHz (See Figure S6), that is, electrical conduction (not ionic). Since the resistivity is very low, we tried to measure temperature-dependent conductivities of this fascinating material in order to estimate activation energy for the conduction. However, the resistivity increased after the measurements up to 1.5 Ωm at 20°C (note the value mentioned above is around the lowest value), and thus we considered that the conductivity is not stable. After these measurements, by using an electrometer, we found that this conductivity is not from the bulk red material but from surface shiny black material. Thus, we speculate that the surface defects worked as carriers, though further details on the origin of the conductivity are under investigation.

**Conclusions**

In summary, a dark red bismuth phase, $(NH_4)_3Bi_2I_9$, with a 2D layered perovskite-like architecture was synthesised in solution and characterised by X-ray diffraction. DFT calculations were in good agreement with the experimental results on the structural solutions. Band gap of the material was measured as 2.04 eV (1.42 eV in DFT calculation). The results of the conductivity measurement suggest that the bulk material is insulating, however, the conductivity in the powder form might be due to the surface effects from making the pallets. X-ray photoelectron spectroscopy and thin film fabrications are under investigation to provide better understanding of the application of this bismuth based perovskite family in photovoltaic devices.

[note] over the detection limit of 1 TΩ as found in previous works.[37,39]

**Acknowledgements**

This work was supported by Cambridge Overseas Trust, Winton Programme for the Physics of Sustainability at the University of Cambridge, Advanced Investigator Award to AKC from the European Research Council (ERC) and the World Premier International Research Center Initiative on "Materials Nanoarchitectonics (WPI-MANA)" from MEXT, Japan. We thank Alan Dickerson (Department of Chemistry, Cambridge) for CHN analysis and Dr. Gregor Kieslich (Department of Materials Science and Metallurgy) for structural solutions.

# Supporting Information

## Synthesis, Crystal Structure and Properties of a Perovskite-Related Bismuth Phase, $(NH_4)_3Bi_2I_9$

Shijing Sun,[a] Satoshi Tominaka,[b] Jung-Hoon Lee,[a] Fei Xie,[a] Paul D. Bristowe[a] and Anthony K. Cheetham[a*]

[a]Functional Inorganic and Hybrid Materials Group, Department of Materials Science and Metallurgy, University of Cambridge, U.K.
[b]National Institute for Materials Science, Tsukuba, Japan

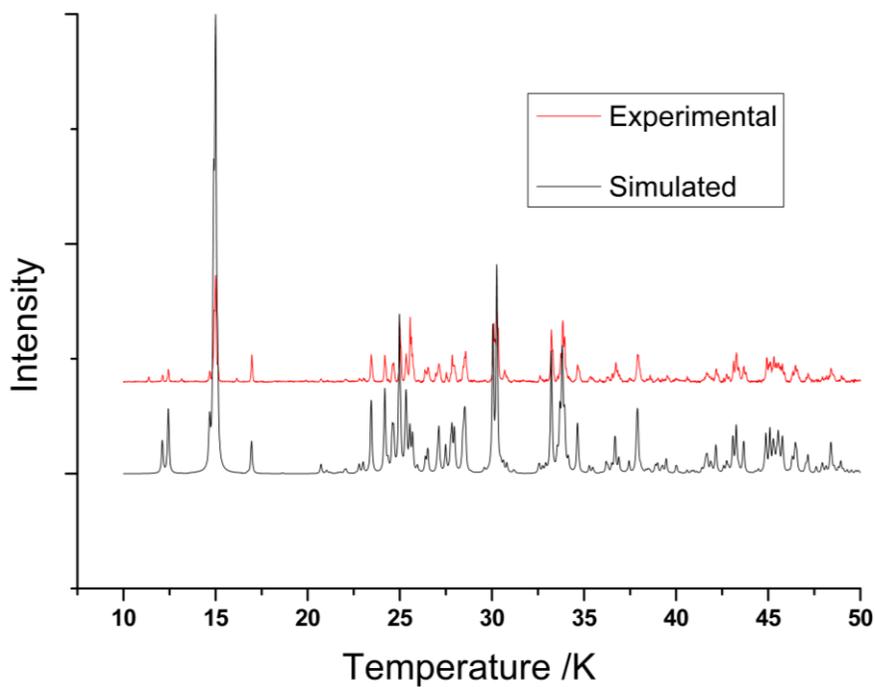

**Figure S1** Powder X-ray diffraction pattern of experimental and simulated $(NH_4)_3Bi_2I_9$ phase.

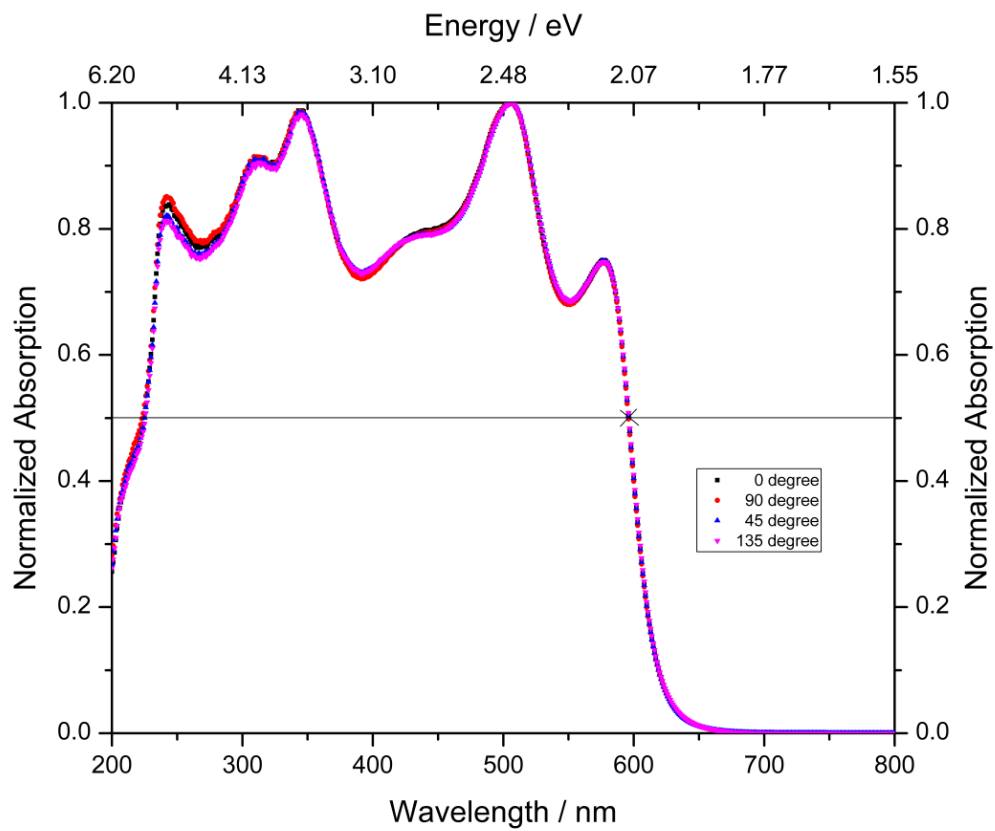

**Figure S2** Absorption spectrum plotted using the reflectance data on powder sample of $(NH_4)_3Bi_2I_9$, the sample was rotated 4 times to ensure accuracy and the average was used to calculate band gap using Tauc relation.

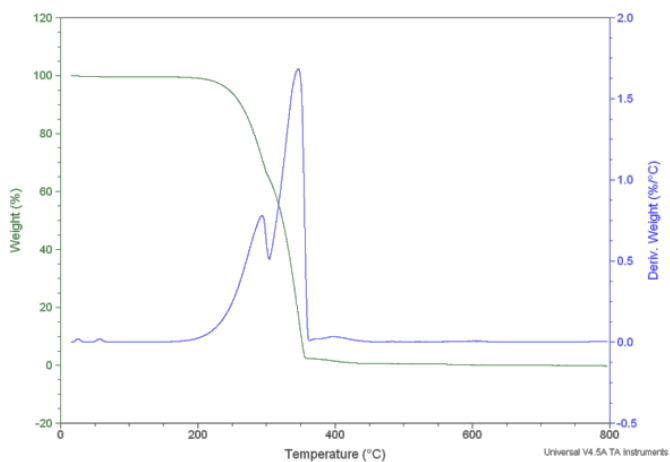

**Figure S3** TGA Thermogravimetric analysis (TGA) curve of $(NH_4)_3Bi_2I_9$ under nitrogen flow.

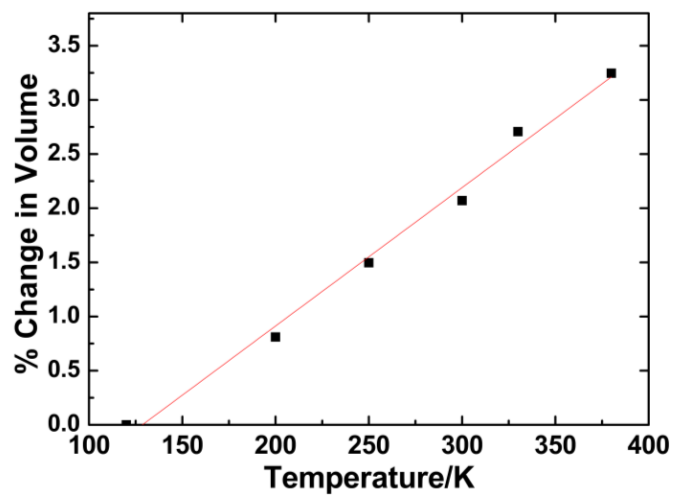

**Figure S4** Change in unit cell volume in the temperature range of 120 K – 380 K.

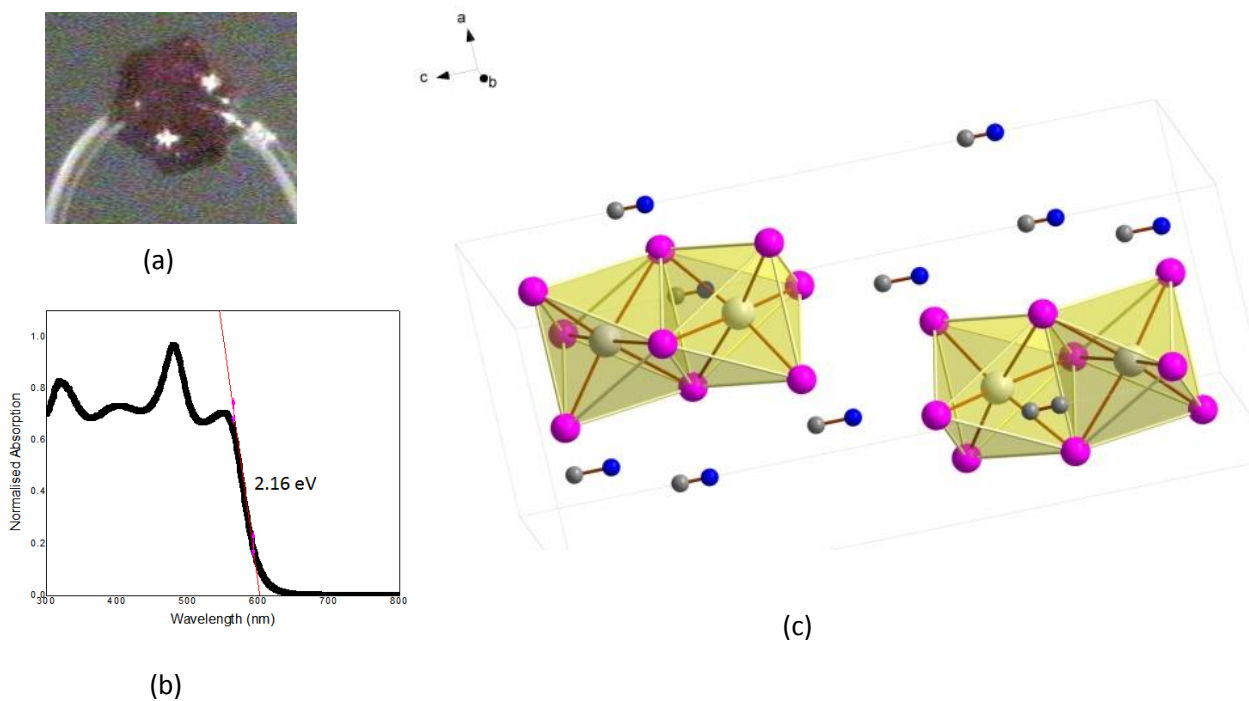

**Figure S5** The dimer structure of (CH$_3$NH$_3$)$_3$Bi2I9, (a) a representative photograph during single crystal X-ray diffraction measurement. (b) Absorption spectrum on powder sample to calculate the band gap. (c) Crystal structure showing the face sharing polyhedral. Silver spheres represent bismuth, purple represent iodine, blue represent Carbon and grey represent nitrogen atoms.

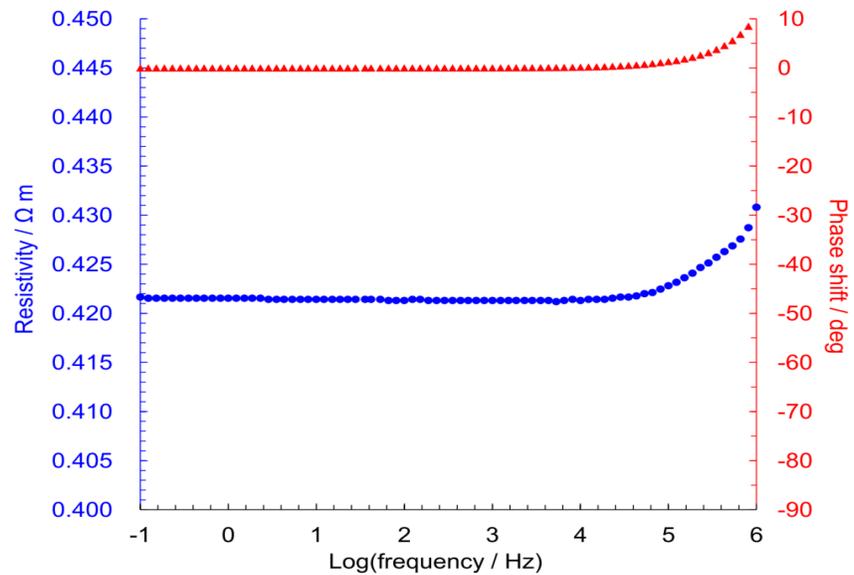

**Figure S6** AC impedance data for the pellet sample (20°C).

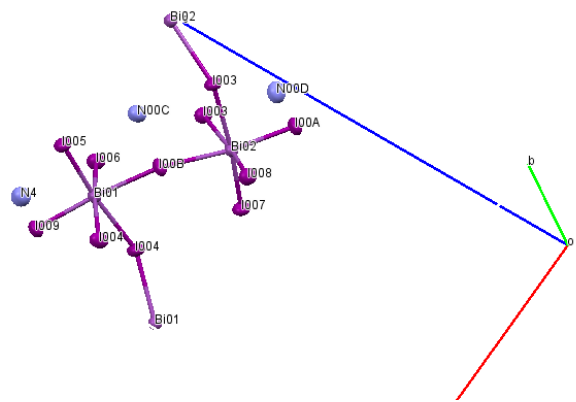

**Figure S7** Bonding observed in (NH$_4$)$_3$Bi$_2$I$_9$ from single crystal X-ray diffraction.

**Table S1** Coefficients of thermal expansion (along *a*, *b*, *c* axis and volumetric) over 120 K - 380 K.

| $\alpha_a/K^{-1}$ | $\alpha_b/K^{-1}$ | $\alpha_c/K^{-1}$ | $\alpha_v/K^{-1}$ |
|---|---|---|---|
| 4.60E-05 | 5.64E-05 | 2.06E-05 | 1.25E-04 |

**Table S2** Crystallographic data of the dimer phase, $(CH_3NH_3)_3Bi_2I_9$.

| Empirical formula | $C_3N_3H_{18}Bi_2I_9$ |
|---|---|
| Temperature/K | 300 |
| Crystal system | Hexagonal |
| Space group | P 6$_3$/m m c |
| a/Å | 8.581 |
| b/Å | 8.581 |
| c/Å | 21.760 |
| α/° | 90 |
| β/° | 90 |
| γ/° | 120 |
| Volume/Å$^3$ | 1387.59 |

**Table S3** Experimental fractional atomic coordinates (×10$^4$) and equivalent isotropic displacement parameters (Å$^2$×10$^3$), following the atom labels in Figure S7. U$_{eq}$ is defined as 1/3 of of the trace of the orthogonalised U$_{IJ}$ tensor.

| Atom | x | y | z | U(eq) |
|---|---|---|---|---|
| Bi01 | 6669.7(2) | 4940.9(4) | 8441.6(2) | 37.46(12) |
| Bi02 | 3360.2(2) | 4891.8(4) | 6542.8(2) | 37.73(12) |
| I003 | 2396.8(5) | 8125.1(8) | 7162.9(4) | 51.5(2) |
| I004 | 6967.3(4) | 1760.9(8) | 7547.2(4) | 53.6(2) |
| I005 | 6258.6(5) | 7972.3(8) | 9152.0(4) | 54.6(2) |
| I006 | 5428.9(5) | 2970.5(8) | 9225.0(4) | 54.6(2) |
| I007 | 4296.4(5) | 1897.2(8) | 6128.0(4) | 54.8(2) |
| I008 | 4125.0(5) | 6934.6(8) | 5528.9(4) | 55.2(2) |
| I009 | 8305.5(4) | 4185.3(9) | 9226.8(5) | 59.7(2) |
| I00A | 1665.4(5) | 4226.5(10) | 5806.6(5) | 63.4(3) |
| I00B | 5068.4(5) | 5956.6(10) | 7442.1(5) | 66.0(3) |
| N00C | 4659(6) | 10266(11) | 7780(6) | 63(3) |
| N00D | 1901(7) | 9763(12) | 5601(7) | 76(3) |
| N4 | 8530(6) | 9730(12) | 8920(6) | 74(3) |

**Table S4** Experimental bond lengths of $(NH_4)_3Bi_2I_9$, following the atom labels in Figure S7.

| Atom | Atom | Length/Å | Atom | Atom | Length/Å |
|---|---|---|---|---|---|
| Bi01 | I004[1] | 3.2508(8) | Bi02 | I003[2] | 3.2729(9) |
| Bi01 | I004 | 3.2275(8) | Bi02 | I007 | 2.9337(7) |
| Bi01 | I005 | 2.9476(8) | Bi02 | I008 | 2.9300(9) |
| Bi01 | I006 | 2.9392(8) | Bi02 | I00A | 2.9448(8) |
| Bi01 | I009 | 2.9428(8) | Bi02 | I00B | 3.2195(8) |
| Bi01 | I00B | 3.2198(9) | I003 | Bi02[3] | 3.2730(9) |
| Bi02 | I003 | 3.2637(7) | I004 | Bi01[4] | 3.2508(8) |

[1]3/2-X,1/2+Y,3/2-Z; [2]1/2-X,-1/2+Y,3/2-Z; [3]1/2-X,1/2+Y,3/2-Z; [4]3/2-X,-1/2+Y,3/2-Z

**Table S5** Bond Angles of $(NH_4)_3Bi_2I_9$, following the atom labels in Figure S7.

| Atom | Atom | Atom | Angle/° | Atom | Atom | Atom | Angle/° |
|---|---|---|---|---|---|---|---|
| I004 | Bi01 | I004[1] | 84.701(12) | I007 | Bi02 | I003[2] | 92.54(2) |
| I005 | Bi01 | I004 | 173.85(3) | I007 | Bi02 | I00A | 94.86(2) |
| I005 | Bi01 | I004[1] | 94.05(2) | I007 | Bi02 | I00B | 92.05(2) |
| I005 | Bi01 | I00B | 87.72(2) | I008 | Bi02 | I003 | 90.08(2) |
| I006 | Bi01 | I004[1] | 173.15(2) | I008 | Bi02 | I003[2] | 170.26(2) |
| I006 | Bi01 | I004 | 88.57(2) | I008 | Bi02 | I007 | 94.19(2) |
| I006 | Bi01 | I005 | 92.51(2) | I008 | Bi02 | I00A | 93.06(3) |
| I006 | Bi01 | I009 | 94.44(3) | I008 | Bi02 | I00B | 88.21(3) |
| I006 | Bi01 | I00B | 93.09(2) | I00A | Bi02 | I003 | 89.53(2) |
| I009 | Bi01 | I004 | 92.28(2) | I00A | Bi02 | I003[2] | 93.38(2) |
| I009 | Bi01 | I004[1] | 87.13(2) | I00A | Bi02 | I00B | 172.86(3) |
| I009 | Bi01 | I005 | 93.67(3) | I00B | Bi02 | I003 | 83.45(2) |
| I009 | Bi01 | I00B | 172.27(3) | I00B | Bi02 | I003[2] | 84.52(3) |
| I00B | Bi01 | I004[1] | 85.19(2) | Bi02 | I003 | Bi02[3] | 146.79(3) |
| I00B | Bi01 | I004 | 86.17(2) | Bi01 | I004 | Bi01[4] | 145.47(2) |
| I003 | Bi02 | I003[2] | 82.674(12) | Bi02 | I00B | Bi01 | 149.12(3) |
| I007 | Bi02 | I003 | 173.71(3) | | | | |

[1]3/2-X,1/2+Y,3/2-Z; [2]1/2-X,-1/2+Y,3/2-Z; [3]1/2-X,1/2+Y,3/2-Z; [4]3/2-X,-1/2+Y,3/2-Z